\documentclass[a4]{article}
\usepackage{a4wide}
\usepackage[centertags]{amsmath}

\def\1ad{\mbox{\normalsize $^1$}}
\def\2ad{\mbox{\normalsize $^2$}}
\def\3ad{\mbox{\normalsize $^3$}}
\def\4ad{\mbox{\normalsize $^4$}}
\def\5ad{\mbox{\normalsize $^5$}}
\def\6ad{\mbox{\normalsize $^6$}}
\def\7ad{\mbox{\normalsize $^7$}}
\def\8ad{\mbox{\normalsize $^8$}}
\def\makefront{\vspace*{1cm}\begin{center}
\def\newtitleline{\\ \vskip 5pt}
{\Large\bf\titleline}\\
\vskip 1truecm
{\large\bf\authors}\\
\vskip 5truemm
\addresses
\end{center}
\vskip 1truecm
{\bf Abstract:}
\abstracttext
\vskip 1truecm}

\let\rho\varrho
\let\phi\varphi

\newcommand{\Spin}{\text{Spin}} 
\newcommand{\pSU}{\text{SU}}    
\newcommand{\pU}{\text{U}  }     
\newcommand{\pE}{\text{E}  }     

\newcommand{\mR}{{\bf R}}
\newcommand{\mC}{{\bf C}}
\newcommand{\mZ}{{\bf Z}}

\newcommand{\auf}{\rightarrow}

\DeclareMathOperator{\tr}{tr}

\newcommand{\qed}{\ \hfill $\Box$\par}

\newcommand\DDS{\text{\hbox to 0pt{/\hss}D}}
\newcommand\DpS{\text{\hbox to 0pt{/\hss}}\partial}

\begin {document}
\def\titleline{
  On Fractional Instanton Numbers in Six Dimensional
  \newtitleline
  Heterotic $\pE_8 \times \pE_8$ Orbifolds
}
\def\authors{
Jan O. Conrad
}
\def\addresses{
Physikalisches Institut, Universit\"at Bonn\\
Nu\ss{}allee 12, D-53115 Bonn, Germany
}
\def\abstracttext{
  We show how the level matching condition in six dimensional, abelian
  and supersymmetric orbifolds of the $\pE_8 \times \pE_8$ heterotic
  string can be given equivalently in terms of fractional gauge and
  gravitational instanton numbers. This relation is used to restate
  the classification of the orbifolds in terms of flat bundles away
  from the orbifold singularities under the constraint of the level
  matching condition. In an outlook these results are applied to
  Kaluza-Klein monopoles of the heterotic string on $S^1$ in Wilson
  line backgrounds.
}
\large
\makefront

\section{Introduction}

In this talk we present some results of \cite{Conrad} and apply them
to Kaluza-Klein monopoles of the heterotic string on $S^1$.

Even though heterotic orbifolds have been known for a long time
\cite{OrbI, OrbII}, their strong coupling description in terms of
M-theory on $S^1/\mZ_2$ is still very limited \cite{Stieberger,
  OvrutI, Kaplunovsky, OvrutII, OvrutIII, OvrutIV}.  Here we focus on
symmetric, abelian and perturbative orbifolds of the $\pE_8 \times
\pE_8$ heterotic string in six dimensions preserving 8 supersymmetries
without discrete torsion.

\section{The level matching condition}
\label{one}

At first, we consider a single fixed point in an arbitrary model
located at the origin of $\mC^2$. Associated to it is a generator
$(r,\gamma)$ of $\mZ_N$ consisting of a rotation $r$ acting like
$\exp(2\pi i \Phi_i)$ on the complex coordinate $Z^i$ and a gauge
shift $\gamma$ acting like $\exp(2\pi i \beta^I_1 H^1_I)$ $\exp(2\pi i
\beta_2^I H^2_I)$ where the $H_I^{1,2}$ are the eight generators of
the Cartan subalgebra in the adjoint representation of
$\pE_8^{(1,2)}$. The $H_I$ will be normalized such that the $\pE_8$
lattice $\Lambda_8$ is given by the vectors $p^I = (n_1, \ldots, n_8)$
and $p^I = (\tfrac{1}{2} + n_1, \ldots, \tfrac{1}{2} + n_8)$ with $n_i
\in \mZ$ and $\sum_i n_i \in 2\mZ$. This implies that $q^I_{1,2} =
N\beta^I_{1,2}$ is a lattice vector.

Consistency of the model requires the level matching condition
\begin{equation}
  \label{eq:level}
  \Phi^2 = \beta_1^2 + \beta_2^2 \mod \frac{2}{N}
\end{equation}
to be satisfied \cite{OrbII,Vafa:1986wx}. This equation, however,
bears great similarity to the relation
\begin{equation}
  \label{eq:GreenSchwarz}
  \tr R^2 = \frac{1}{30} \tr F_1^2 + \frac{1}{30} \tr F_2^2
\end{equation}
required by the Green-Schwarz mechanism \cite{GreenSchwarz} together
with perturbativity of the orbifold. Therefore one is naturally led to
the question of to what extent the two equations are related. But
already in the case of the simplest $\mZ_3$ orbifolds it can be seen
that \eqref{eq:level} can not account for the full instanton numbers.
To this end, consider the standard embedding $(1,-1,0^6;0^8)$ with the
gauge group $\pU(1) \times \pE_7 \times \pE_8$ and compare to the
embedding $(1,-1,0^6;2,1^2,0^5)$ with the gauge group $\pU(1) \times
\pE_7 \times \pSU(3) \times \pE_6$. In the first case the instanton
number for the second $\pE_8$ is zero whereas in the second case it
cannot be zero!

However, by comparing to the $\Spin(32)/\mZ_2$ case
\cite{Berkooz,Intriligator} one can easily guess \cite{Aldazabal} that
the level matching condition \eqref{eq:level} corresponds to the
formula
\begin{equation}
  \label{eq:level2}
  - \frac{1}{2} \frac{1}{8\pi^2} \int_U \tr R^2 =
  - \frac{1}{60} \frac{1}{8\pi^2} \int_U \tr F_1^2
  - \frac{1}{60} \frac{1}{8\pi^2} \int_U \tr F_2^2
  \pmod 1
\end{equation}
As was shown in \cite{Conrad}, the fractional part of the instanton
number for $\pE_8$ is indeed given by
\begin{equation}
  \label{eq:Main}
  I = - \frac{1}{60} \frac{1}{8\pi^2} \int_U \tr F^2
  = \frac{N}{2} \beta^2
  \pmod 1
\end{equation}
Note that this formula is clearly invariant under the Weyl group as
well as under lattice shifts $\beta \mapsto \beta + p$, since
\begin{equation}
  I
  = \frac{N}{2} (\beta + p)^2
  = \frac{N}{2} (\beta^2 + 2pq/N + p^2)
  = \frac{N}{2} \beta^2 + pq + N \frac{p^2}{2}
  = \frac{N}{2} \beta^2 \pmod 1
\end{equation}
and $pq \in \mZ$, $p^2 \in 2\mZ$.

Therefore the level matching condition in the form of
\eqref{eq:level2} corresponds to the condition that at every fixed
point the sum of the fractional parts of the gauge instanton numbers
match the fractional part of the gravitational instanton number (which
is $1/N$).  In particular the level matching condition as given in
\eqref{eq:level} has nothing to say on how the integer part of the
gauge instanton numbers gets distributed among the two $\pE_8$ groups,
i.e. the two walls in M-theory on $S^1/\mZ_2$.

\section{Global description of the orbifold}

We now turn to the global description of the orbifold. The target
space, i.e. the geometric orbifold, is conveniently described as $O =
\mC^2/S$ where $S$, the space group, is generated by affine
transformations $D : x \mapsto rx + h$ \cite{OrbII} (for an
introduction see \cite{Nilles} or \cite{Pol}, chapter 16). As we
restrict to abelian orbifolds, the point group, i.e. the group
generated by the rotations $r$ alone, is abelian.

To include the gauge degrees of freedom we augment $D$ by a gauge
transformation $\gamma_D$ and require that the map $\gamma : S \auf
\pE_8 \times \pE_8$ is a group homomorphism. This implies that, when a
nontrivial group element is associated to a pure translation $h$, it
generates a cyclic subgroup of $\pE_8$. In this case the model is said
to contain ``discrete Wilson lines''.

As we will now show, the map $\gamma$ is nothing but the data of a
flat $\pE_8 \times \pE_8$ bundle on the orbifold with the fixed points
taken out. Define $F$ to be the set of fixed points of elements of
$S$.  Then the fundamental group of $\mC^2 - F$ is trivial since, by
supersymmetry, $F$ consists of isolated points only.  Therefore $\mC^2
- F$ is the universal covering space of $(\mC^2-F)/S = O-F$ and the
fundamental group of $O-F$ is given by $S$. This should be intuitively
clear, since every line in $\mC^2 - F$ whose ends are identified under
an nontrivial element of $S$, gives rise to a nontrivial loop in
$O-F$.  But since a flat bundle, with respect to a fixed point of
reference, is given by the gauge transformations associated to every
nontrivial loop starting and ending at the reference point, the map
$\gamma$ specifies a flat bundle and vice versa (up to gauge
transformations of the bundle).

Consistency of the model requires that for each fixed point $x_0 =
Dx_0$ with $Dx = rx + h$ the level matching condition \eqref{eq:level}
has to be fulfilled for the generator $(r, \gamma_D)$. This translates
into the requirement that the flat bundle gives rise to fractional
instanton numbers satisfying \eqref{eq:level2} at every fixed point.

\section{Conclusions}

We have shown that the orbifolds in our class correspond to all
possible flat $\pE_8 \times \pE_8$ bundles on the orbifold $\mC^2/S$
with the fixed points taken out under the only restriction that the
sum of the fractional parts of the gauge instanton numbers (computed
from the flat bundle data via \eqref{eq:Main}) match the fractional
part of the gravitational one locally at every fixed point.

Since the fractional instanton numbers are computed seperately for
every $\pE_8$, this classification fully applies to M-theory on
$S^1/\mZ_2$.

{}Furthermore, as there exists a model for any given data satisfying the
level matching conditions, these conditions should account for all
global consistency conditions, at least from the ten dimensional
viewpoint. This implies, by locality, that anomaly cancellation in
M-theory on $S^1/\mZ_2$ can be (and must be) fully accounted for by
discussing single fixed points as in section~\ref{one}.

\section{Outlook}

We end with a small outlook at possible uses of the computation given
in \cite{Conrad}.

One of the motivations for \cite{Conrad} was that the gauge shift at a
given fixed point is given by a vector $\beta$ which is right the
amount of data needed to specify a Wilson line, say of the heterotic
string on $S^1$. To make this similarity more explicit, consider an
open ball cut out around the origin of $\mC^2$. Dividing by $\mZ_N$ as
in section~\ref{one} we get an $A_{N-1}$ orbifold singularity at the
origin, which can be easily blown up to a manifold $M$ with boundary
$S^3/\mZ_N =: L_N$. This boundary is known as a lens space.

It is well known that this configuration can be found at the core of
$N$ Kaluza-Klein monopoles of the heterotic string on $S^1$ sitting
nearly on top of each other \cite{Sen}. Furthermore, taking an $S^2$
surrounding the monopoles in nine dimensions this sphere lifts to a
bundle $S^1 \auf S^2$ in ten dimensions. But the total space of that
bundle is nothing but our lens space $L_N$: in case of one KK-monopole
the $S^1$ bundle is well known to be the Hopf-fibration which is
nothing but an $\pU(1)$ bundle on $S^2$ with transition function $1
\in \pi_1(\pU(1))$. Combining $N$ monopoles just gives an $\pU(1)$
bundle with transition function $N \in \pi_1(\pU(1))$ which can be
identified with $L_N$ (see \cite{Conrad}).

But very far from the centers of the KK-monopoles spacetime looks
locally like $S^1 \times \mR^3 \times \mR^{5,1}$ and the $Z_N$
generator just rotates one time around the $S^1$. Therefore,
associating a gauge shift $\beta$ with this generator is equivalent to
specifying a Wilson line given by $\beta$ around the $S^1$.

Now one of the features of the calculation of \eqref{eq:Main} given in
\cite{Conrad} was that for a gauge field configuration where the gauge
connection $A$ is constant around the $S^1$, i.e. a Wilson line, even
the integer part of the instanton number can be computed:
\begin{equation}
  I = - \frac{1}{60} \frac{1}{8\pi^2} \int_U \tr F^2
  = \frac{N}{2} \beta^2
\end{equation}

Even though such a configuration cannot be supersymmetric because of
the sign of the instanton number, it should appear in the magnetic
charge lattice of the heterotic string on $S^1$.

This correspondence might shed some light on the physics of the
heterotic string on orbifold singularities with nontrivial gauge
backgrounds.

\vskip0.5cm
\noindent
{\large \bf Acknowledgements}

\smallskip
\noindent
This work was supported by the European Programs HPRN-CT-2000-00131,
HPRN-CT-2000-00148 and HPRN-CT-2000-00152 and by the DFG
Schwerpunktprogramm (1096). I would like to thank A.~Font, N.~Obers,
S.~Theisen and A.~Uranga for discussion.


\end{document}